\documentclass[10pt, a4paper]{article}
\usepackage{INTERSPEECH2020}
\usepackage{amsmath,graphicx}
\usepackage{mathrsfs}
\usepackage{amsfonts}
\usepackage{color, soul}
\usepackage{ctable}
\usepackage{multirow}
\usepackage{xspace}

\newcolumntype{P}[1]{>{\centering\arraybackslash}p{#1}}

\makeatletter
\DeclareRobustCommand\onedot{\futurelet\@let@token\@onedot}
\def\@onedot{\ifx\@let@token.\else.\null\fi\xspace}

\def\eg{\emph{e.g}\onedot,\ } 
\def\ie{\emph{i.e}\onedot,\ } 
 
\def\etc{\emph{etc}\onedot}

\makeatother

\title{Speaker attribution with voice profiles by graph-based \\ semi-supervised learning}

\name{Jixuan Wang$^{1,2*}$\thanks{*Work done by the first author during internship at Microsoft.}, Xiong Xiao$^{3}$, Jian Wu$^{3}$, Ranjani Ramamurthy$^{3}$,\\ Frank Rudzicz$^{1,2}$, Michael Brudno$^{1,2}$}
\address{$^{1}$University of Toronto, Canada \\
$^{2}$Vector Institute, Canada \\
$^{3}$Microsoft, USA}

\email{$^{1,2}$\{jixuan, frank, brudno\}@cs.toronto.edu, $^{3}$\{xioxiao, jianwu, ranjanir\}@microsoft.com}

\begin{document}
\maketitle

%%%%%%%%%%%%%%%%%%%%%%%%%%%%%%%%%%%%%%%%%%%%%%%%%%
% Abstract
%%%%%%%%%%%%%%%%%%%%%%%%%%%%%%%%%%%%%%%%%%%%%%%%%%
\begin{abstract}
Speaker attribution is required in many real-world applications, such as meeting transcription, where speaker identity is assigned to each utterance according to speaker voice profiles. In this paper, we propose to solve the  speaker attribution problem by using graph-based semi-supervised learning methods. A graph of speech segments is built for each session, on which segments from voice profiles are represented by labeled nodes while segments from test utterances are unlabeled nodes. The weight of edges between nodes is evaluated by the similarities between the pretrained speaker embeddings of speech segments. Speaker attribution then becomes a semi-supervised learning problem on graphs, on which two graph-based methods are applied: label propagation (LP) and graph neural networks (GNNs). The proposed approaches are able to utilize the structural information of the graph to improve speaker attribution performance. Experimental results on real meeting data show that the graph based approaches reduce speaker attribution error by up to 68\% compared to a baseline speaker identification approach that processes each utterance independently.
\end{abstract}
\noindent\textbf{Index Terms}: speaker attribution, speaker identification, graph neural networks, label propagation.

%%%%%%%%%%%%%%%%%%%%%%%%%%%%%%%%%%%%%%%%%%%%%%%%%%
% Introduction
%%%%%%%%%%%%%%%%%%%%%%%%%%%%%%%%%%%%%%%%%%%%%%%%%%
\section{Introduction}
\label{sec:intro}
Speaker diarization is the problem of ``who spoke when", \ie{grouping} the segments of a long audio recording into speaker-homogeneous clusters. The conventional speaker diarization task assumes no prior knowledge of speakers' identities, so it is basically a clustering problem without speaker identification. However, there are scenarios, such as meeting transcription, where voice profiles of speakers are available and the identification of speakers is required. This task is called `speaker attribution' in this paper. A straightforward approach is to build a multi-class classifier from the speaker profiles, and then classify the test segments one-by-one. A drawback of this approach is treating test segments independently without considering the context when making predictions. For example, those test segments that are similar to each other should be assigned to the same speaker.

\begin{figure}[t]
\begin{minipage}[b]{.48\linewidth}
  \centering
  \centerline{\includegraphics[width=3.8cm]{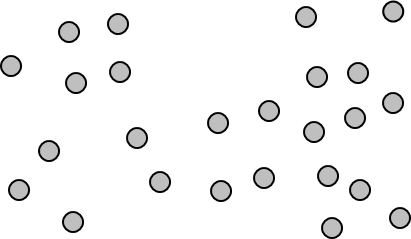}}
%  \vspace{2.0cm}
  \centerline{(a) Extract d-vectors}\medskip
\end{minipage}
\begin{minipage}[b]{.48\linewidth}
  \centering
  \centerline{\includegraphics[width=3.8cm]{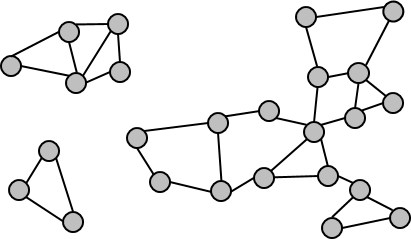}}
%  \vspace{2.0cm}
  \centerline{(b) Build a graph}\medskip
\end{minipage}
\begin{minipage}[b]{.48\linewidth}
  \centering
  \centerline{\includegraphics[width=3.8cm]{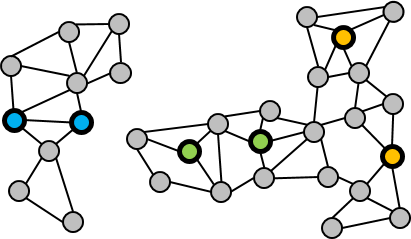}}
%  \vspace{1.5cm}
  \centerline{(c) Add profile segments}\medskip
  \label{fig:semi-graph}
\end{minipage}
\hfill
\begin{minipage}[b]{0.48\linewidth}
  \centering
  \centerline{\includegraphics[width=3.8cm]{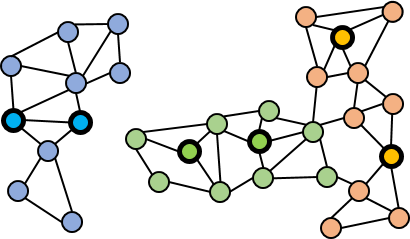}}
%  \vspace{1.5cm}
  \centerline{(d) Label prediction}\medskip
\end{minipage}
\caption{Overview of the proposed method: (a) extract d-vectors of audio segments with a pre-trained speaker embedding model; (b, c) build a graph of audio segments based on pair-wise similarities of the corresponding d-vectors, using both profile and test audio segments; (d) predict labels for test audio segments by graph-based semi-supervised learning methods.}
\label{fig:overview}
\end{figure}

% and hence is theoretically suboptimal. 
Instead of predicting the speaker label for each speech segment independently, we propose to use graph-based semi-supervised learning methods that use the structural information among speech segments within a session. Each speech segment, either from profile audio or test audio, is represented as a node on a graph for each session.
The feature of each node is represented by a fixed-dimensional speaker embedding, \eg{d-vectors}~\cite{variani2014deep, snyder2018x, li2017deep, bredin2017tristounet,wan2018generalized, wang2019centroid, zhou2019cnn}, extracted from the corresponding speech segment.
% Thanks to the recent progress in deep learning based speaker embedding, we can represent the speaker characteristics of a speech segment as a fixed-dimensional vector no matter how long the segment is.
Segments from the profile audio are treated as labeled nodes while those from the test audio are unlabeled nodes.
The speaker attribution task can then be solved as a graph-based semi-supervised learning problem, which can now utilize the structural information of the graph in order to improve the accuracy of classifying the test nodes. The intuition is that if two nodes are similar to each other and share common neighbors on the graph, they are likely to have the same speaker label. Recently, graph-based methods have also been successfully applied on the conventional speaker diarization problem~\cite{jixuan2020speaker}.

An overview of the proposed method is shown in Figure~\ref{fig:overview}. 
First, we apply a pre-trained speaker embedding model to extract d-vectors for speech segments, which are obtained by uniformly segmenting the audio of one session after applying voice activity detection (VAD). Then a graph is built with both speech segments from profile audio and test audio as shown in Figure~\ref{fig:overview}(c), on which each node is a speech segment whose average d-vector is used as the feature vector of the node. The weight of each edge represents the similarity between the corresponding segment pair. The profile segments are labeled nodes on the graph, while test segments are unlabeled. To classify the unlabeled nodes, we apply two graph-based semi-supervised learning methods: a graph Laplacian regularization-based approach (label propagation) and a graph embedding-based approach by GNNs. Experiments show that both of these two methods significantly outperform the classification-based methods on real multi-party meetings and present great potential for real-world applications.  
Our contributions can be summarized as:
\begin{itemize}
    \item we propose the first solution to speaker attribution with speaker profiles through graph-based semi-supervised learning methods;
    \item we study two graph-based methods -- label propagation and GNN-based  -- and their applications on a speaker attribution pipeline; and
    \item we evaluate the proposed methods on real meeting data. Results show that the graph-based methods significantly outperform the baseline method and present great potential for real-world applications.
\end{itemize}

%%%%%%%%%%%%%%%%%%%%%%%%%%%%%%%%%%%%%%%%%%%%%%%%%%
% Related Work
%%%%%%%%%%%%%%%%%%%%%%%%%%%%%%%%%%%%%%%%%%%%%%%%%%
\section{Related work}
\label{sec:related}

\subsection{Speaker diarization with speaker profiles}
While the conventional speaker diarization problem is usually solved by clustering or end-to-end approaches~\cite{meignier2010lium, shum2013unsupervised, sell2014speaker, garcia2017speaker, wang2018speaker, zhang2019fully}, speaker diarization with profiles can be handled as a speaker identification or classification task. Speech embedding models that map raw speech features into a low-dimensional space are widely used for speaker identification or classification. Traditional methods apply probabilistic linear discriminant analysis (PLDA) on top of i-vectors to classify speakers~\cite{dehak2010front}.
Recently, neural network-based speaker embedding models have become popular in which speaker embedding models are usually trained via classification loss~\cite{variani2014deep, snyder2018x}, triplet loss~\cite{li2017deep, bredin2017tristounet}, generalized end-to-end loss~\cite{wan2018generalized}, or prototypical network loss~\cite{wang2019centroid}.
%Different types of classifier can be built upon the d-vectors and have showed promising results.
The similarities between resulting speaker embeddings can be measured by simple metrics, such as cosine or Euclidean distances, and
a speech segment is typically classified as the speaker whose profile embedding is closest to it in the embedding space.

\subsection{Graph based semi-supervised learning}
Graph-based semi-supervised classification methods perform classification on graphs, where a small fraction of the nodes are labeled. The label information is propagated from the labeled nodes to the unlabeled nodes along the structure of a graph. This can be formulated as adding a graph-based regularization term to the supervised loss~\cite{zhu2003semi, zhou2004learning, belkin2006manifold, weston2012deep}.
%Moreover, graph embedding-based methods have made significant progress recently.
Instead of using explicit graph-based regularization, other types of methods utilize graph neural networks (GNNs) to encode the graph structure and learn new representations for both labeled and unlabeled nodes~\cite{wu2019comprehensive, cai2018comprehensive, kipf2016semi, gilmer2017neural}. The supervised loss is defined on the labeled nodes but the gradient can be distributed to the unlabeled nodes to learn new representations for all nodes, upon which a classifier can be trained for inference.

%%%%%%%%%%%%%%%%%%%%%%%%%%%%%%%%%%%%%%%%%%%%%%%%%%
% Graph based methods for SD
%%%%%%%%%%%%%%%%%%%%%%%%%%%%%%%%%%%%%%%%%%%%%%%%%%
\section{Speaker classification on graphs}
\label{sec:graph}
In this section, we discuss graph-based methods for speaker attribution with speaker profiles. The approach is illustrated in Figure~\ref{fig:overview}.

\subsection{Building similarity graphs of speech segments}
We build a graph for the audio segments of each meeting. Each node represents an audio segment, which could be word-level or utterance-level, or be extracted by a sliding window with a fixed window shift. We use the average d-vector of each segment as the node features. The weight of edges between nodes is represented by the cosine similarity of node features, which is normalized to $[0, 1]$ linearly.

There are several methods to construct graphs from pairwise similarities~\cite{von2007tutorial}: (1) simply connect all nodes, and weigh all edges by the similarities between their nodes; (2)  only connect two nodes if at least one node or both are among the $k$-nearest neighbors of the other; (3) only keep the edges on which the weight is larger than a threshold. In this paper, we apply method (3) and treat the threshold as a hyperparameter.

A meeting session, including related speaker profiles, can be represented as a graph $\mathcal{G}(\mathcal{V}, \mathcal{E}, A)$, where $\mathcal{V}$ is the set of nodes (speech segments), $\mathcal{E}$ is the set of edges, and $A \in \mathbb{R}_{\geq0}^{N \times N}$ is the affinity matrix with $A_{ij} > 0$ if edge $e_{ij} = (v_i, v_j) \in \mathcal{E}$ and  $A_{ij} = 0$ otherwise. $N$ is the total number of nodes. $X=[\mathbf{x}_1,...,\mathbf{x}_N] \in \mathbb{R}^{N \times D}$ is the node feature matrix, where $\mathbf{x}_i$ is the average d-vector of the $i^{th}$ node and $D$ is the dimension of the embedding space. Without loss of generality, we assume the first $M$ nodes are from speaker profiles and hence labeled, and $0 < M < N $.
%The goal of speaker diarization can be formed as prediction of $y_i$ for each segment embedding $\mathbf{x}_i$, $i \in {\{L+1, L+2, \dots, N\}}$, such that $y_i$ is equal to $y_j$ if and only if $\mathbf{x}_i$ and $\mathbf{x}_j$ belong to the same speakers, $j \in {\{1, 2, \dots, N\}}$.

\subsection{Label propagation}
For a meeting session, let $\mathcal{F} \in \mathbb{R}^{N \times C}$ be a set of matrices, where $C$ is the number of speaker classes. Given a $F\in\mathcal{F}$,  $y_i = \arg\max_{1 \leq j \leq C} F_{ij}$ is the predicted speaker ID for node $i$. The label propagation algorithm is summarized as follows:
\begin{enumerate}
    \item Construct the affinity matrix $A$ with $A_{ij} = \frac{1 + \cos{(\mathbf{x}_i, \mathbf{x}_j})}{2} $ and $A_{ii} = 0$, where $\cos(\textbf{a},\textbf{b})$ denotes the cosine similarity of $\textbf{a}$ and $\textbf{b}$.
    \item  Initialize $F$ as $F^{(0)}_{ij} = 1$ if $j = l_i, i \leq M$ and $F^{(0)}_{ij} = 0$ otherwise, where $l_i$ is the label of node $i$
    \item Iteratively update  $F$ using $F^{(t + 1)} = \alpha S F^{(t)}+(1-\alpha)F^{(0)}$, where $S=D^{-1/2}AD^{-1/2}$, $D$ is the degree matrix of $A$, $\alpha$ is a hyperparameter between $(0, 1)$ and $t$ is the iteration index.
    \item  Stop when $F$ has converged to $F^*$. The speaker ID of node $i$ is obtained as $y_i = \arg\max_{1 \leq j \leq C}F_{ij}^*$. 
\end{enumerate}
The initial matrix $F^{(0)}$ represents our knowledge about the labels of the nodes. Through iterative updating of $F$, the label information is propagated to the whole graph. In each iteration, the soft label assignment of a node ($F^{(t)}_i$) is updated as a weighted sum of the soft label assignment of itself, its neighbors, and the initial assignment. In this way, each neighborhood in the graph will tend to have the same speaker ID. A limitation of label propagation is that it does not directly minimize the expected speaker classification errors through learning on the graph. It also lacks the free parameters that allows learning from labeled nodes. More details on label propagation can be found in~\cite{zhou2004learning}. 

In this work, we modified the label propagation by freezing the columns of $F$ for labeled nodes, as we don't want to change the label assignment of the profile segments. We found that this improves the robustness of label propagation on the speaker attribution task. We also use a fixed number of iterations instead of waiting until the algorithm converges, as the converged $F$ does not always give the best performance.

\subsection{Graph embedding based methods}

\begin{figure}[!t]
\begin{minipage}[b]{1.0\linewidth}
  \centering
  \centerline{\includegraphics[width=4.5cm]{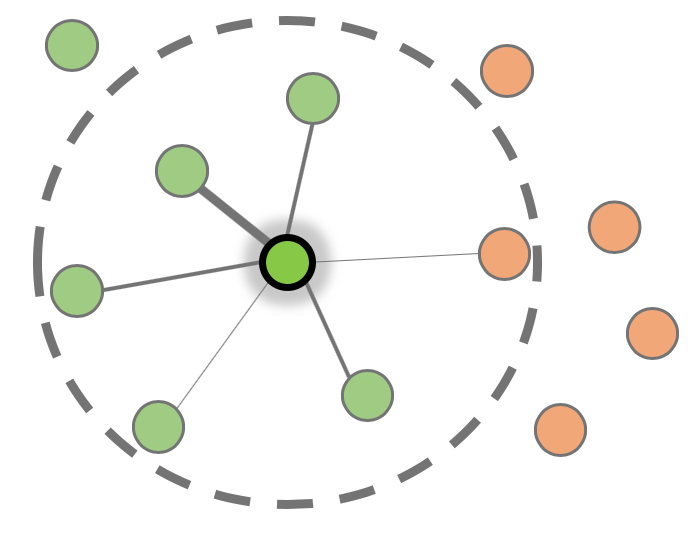}}
\end{minipage}
\caption{Graph convolution. The feature of the bold node in the next layer is the output of a function of the features of itself and its neighbors in the current layer. The thickness of the edges denotes the similarity between pairs of nodes.}
\label{fig:gconv}
\end{figure}

To moderate the limitations of label propagation, we apply several variants of GNNs to the speaker attribution task. Specifically, we apply the variants under the framework of the message passing neural networks (MPNNs)~\cite{gilmer2017neural} and achieve better performance. Under the MPNN framework, as shown in Figure~\ref{fig:gconv}, the convolutional operator is expressed as a message passing scheme:

\begin{equation}
    % \mathbf{x}_i^{(k)} = \gamma^{(k)}_{\mathbf{\Theta}} \left( \mathbf{x}_i^{(k-1)}, \square_{j \in \mathcal{N}(i)} \, \phi^{(k)}\left(\mathbf{x}_i^{(k-1)}, \mathbf{x}_j^{(k-1)},\mathbf{e}_{i,j}\right) \right)
    \mathbf{x}_i^{\prime} = \gamma_{\mathbf{\Theta}} \left( \mathbf{x}_i, \square_{j \in \mathcal{N}(i)} \, \phi_{\mathbf{\Theta}}\left(\mathbf{x}_i, \mathbf{x}_j,\mathbf{e}_{i,j}\right) \right),
\end{equation}
where $\mathbf{x}_i$ is the feature of node $i$ of the current layer with dimension $D$, $\mathbf{x}_i^{\prime}$ is the node feature in the next layer with dimension $D^{\prime}$, $\mathbf{e}_{i,j}$ is the edge feature from node $i$ to node $j$, $\gamma(\cdot)$ and $\phi(\cdot)$ are the update function and message function, respectively (parameterized by $\mathbf{\Theta}$), and $\square$ denotes the aggregation function, \eg{ $sum$, $mean$, $max$}, \etc. $\mathcal{N}{(i)}$ is the set of neighbors of node $i$. 

Specifically, one variant we apply is the graph convolutional network (GCN) described by~\cite{kipf2016semi} in which the $\square$ function corresponds to taking a certain weighted average of neighboring nodes:
\begin{equation}
    \mathbf{x}_i^{\prime} = \sigma \left(W \sum_j L_{ij} \mathbf{x}_j \right),
\end{equation}
where $L = \hat{D}^{-1/2}\hat{A}\hat{D}^{-1/2}$, $\hat{A}=A+I_N$, and $\hat{D}$ denotes the degree matrix of $\hat{A}$. $W \in \mathbb{R}^{D^{\prime} \times D}$ is a layer specific trainable weight matrix, and $\sigma(\cdot)$ is a nonlinear function which, in our case, is the exponential linear units (ELU) activation function~\cite{clevert2015fast}.

We build and train a model with several GCN layers for each meeting session \textit{separately}. The output dimension of the last layer is equal to the number of speaker classes $C$ in a meeting session. Instead of the ELU activation function, we apply the softmax activation function row-wise on the output embedding matrix $X_{out} \in \mathbb{R}^{N \times C}$ from the last GCN layer, resulting in a predicted probability matrix $Z \in \mathbb{R}^{N \times C}$.
The model is trained with regard to the cross-entropy loss over labelled segments in a meeting session:
\begin{equation}
    \mathcal{L} = -\sum_{i=1}^{M} \sum_{j=1}^{C} F_{ij} \ln{Z_{ij}}
\end{equation}

Instead of using explicit graph-based regularization as in label propagation, GNNs can directly encode the graph structure into the model. Training signals can be distributed from the supervised loss $\mathcal{L}$ along the graph structure, and thus the model can learn representations of both labelled and unlabelled speech segments. This enables GNN models to have more powerful classification ability than label propagation. However, for GNN models, we need to separate a subset from the labelled segments in each meeting session for validation purposes.

%%%%%%%%%%%%%%%%%%%%%%%%%%%%%%%%%%%%%%%%%%%%%%%%%%
% Experiments
%%%%%%%%%%%%%%%%%%%%%%%%%%%%%%%%%%%%%%%%%%%%%%%%%%
\section{Experiments}
\subsection{Datasets}
The graph-based methods are evaluated on two sets of in-house real meeting data, including a \textit{dev} set containing 9 meetings and a \textit{test} set containing 19 meetings. The number of speakers in a meeting ranges from 2 to 20, with a mean of 8. The average duration of the meeting sessions is about 41 minutes. The profile audio was recorded by laptop microphones or headphones, and the meeting audio was recorded by microphone arrays in meeting rooms. The duration of the profile audio for a speaker is about 20-40s long and each profile audio file consists of a few sentences. There is no signal processing applied on the profile audio, but the meeting audio is processed by beamforming to enhance the speech signal. After VAD, the profile audio is uniformly split into 1.2s long speech segments, while the meeting audio is split into 0.8s speech segments. For each segment, d-vectors are extracted and averaged to represent the speaker characteristics of the segment. 

%We use the first set of 9 sessions (\textit{dev} set) to search for the optimal hyperparameters and use the remaining 19 sessions (\textit{test} set) for evaluation. 

\subsection{Implementation details}
The d-vector extraction model is trained on the VoxCeleb2 dataset~\cite{chung2018voxceleb2} with data augmentation~\cite{zhou2019cnn}. In our baseline system, each enrolled speaker is represented by a single profile d-vector averaged from the d-vectors extracted from the speaker's profile audio. A segment is assigned to the speaker whose profile d-vector has the highest cosine similarity with the segment's d-vector. 

% The LP method does not require training and only has a few hyperparameters for graph building. 
We use the \textit{dev} set to tune the hyperparameters for graph-based methods, including the threshold for graph edge pruning, parameter $\alpha$ for LP, architecture and learning rate of the GNN model, \etc. 
Specifically, to build graphs we only connect two nodes if their cosine similarity is larger than 0.6.
The GNN model includes two GCN~\cite{kipf2016semi} layers between which we apply the ELU activation function~\cite{clevert2015fast} and a dropout layer~\cite{srivastava2014dropout}. Each hidden layer contains 64 nodes. The output dimension is equal to the number of speakers of a session and a softmax layer is applied to output a probability distribution. The GCN layers are implemented with the PyTorch Geometric library~\cite{fey2019fast}. 

For the GNN-based method, a network is trained for each meeting. We split the labeled nodes into two sets equally and train two models. In the first model, we use the first half of labeled nodes as training set and second half as cross validation set. In the second model, we switch the two sets. In this way, we can effectively use all the labeled nodes for training. For each model, the training stops if there is no improvement on the cross validation loss. During inference, the hidden activation vectors of the two models before the softmax layer are summed. Note the cross validation set here is used for model selection for a single meeting session, which is different from  the \textit{dev} set used for hyperparameter tuning. 

% \begin{figure}[t]
% \begin{minipage}[b]{1.0\linewidth}
%   \centering
%   \centerline{\includegraphics[width=8.0cm]{speaker_number.png}}
% \end{minipage}
% \caption{Speaker number distribution across meeting sessions}
% \label{fig:overview}
% \end{figure}

%%%%%%%%%%%%%%%%%%%%%%%%%
% Experiments design and evaluation metrics
%%%%%%%%%%%%%%%%%%%%%%%%%
\subsection{Experimental design and evaluation metrics}
For each meeting session the same number of profile d-vectors are provided for each speaker.
To evaluate the performance of the methods with different amount of profile data, the number of profile d-vectors provided for training ranged from 5 to 30. The duration of the corresponding profile audio is from 5 to 25 seconds. For each meeting, we apply every method 10 times, each time with randomly sampled profile d-vectors. The profile d-vectors are obtained by randomly sampling consecutive audio segments rather than randomly sampled individual segments in order to be more accordant with practical circumstances. The sampling is performed only once per run, so different methods use exactly the same sampled profile segments. We report both the mean and standard deviation of the segment classification errors across the 10 runs. 

\subsection{Experimental results}
The results on the \textit{dev} and \textit{test} sets are shown in Table~\ref{tbl:validation} and \ref{tbl:test}, respectively. Graph-based methods significantly outperform the baseline method, resulting in much lower error rate and lower standard deviation. The relative error reduction against the baseline method is up to 68.2\% and 48.7\% on the \textit{dev} and \textit{test} sets, respectively. 
The GNN-based method demonstrates better performance over LP under all settings in terms of both accuracy and stability.% We have also tried other variants of GNN models and they result in similar performance.

Although the baseline model is simple, it achieves reasonable results. Applying more complex classifiers, \eg{support vector machines (SVMs) or multilayer perceptron (MLP) models} do not perform better than the baseline due to the small amount of training data (profile data) and the acoustic mismatch between profile audio and meeting audio. The graph-based semi-supervised learning methods were able to learn from both labeled and unlabeled data, alleviating the data sparsity and mismatch problem significantly.

% We also tried a few speaker classification methods such as SVM and MLP, but none of them outperformed the baseline system. The poor performance of SVM and MLP could be due to that there is insufficient training data so that the models are likely to overfit. Also the models have to be re-trained for each meeting since different meetings have different sets of speakers. 

\begin{table}[t]
\caption{Segment error rate on the validation set. ``\#'' denotes the number of labeled speech segments for each speaker. ``Cosine'' refers to the the baseline method based on cosine similarity. ``mean'' and ``std.'' refer to mean value and standard derivation, respectively. ``RER'' refers to relative error reduction with regard to the baseline method.}
\small
\begin{tabular}{c|P{0.4cm}P{0.4cm}|P{0.4cm}P{0.4cm}P{0.8cm}|P{0.4cm}P{0.4cm}P{0.8cm}}
\specialrule{.1em}{.05em}{.05em} 
\multirow{2}{*}{\begin{tabular}[c|]{@{}l@{}}\# \\  
\end{tabular}} & \multicolumn{2}{c|}{Cosine} & \multicolumn{3}{c|}{LP} & \multicolumn{3}{c}{GCN} \\               & mean         & std.        & mean  & std. & RER/\% & mean  & std.  & RER/\% \\ \hline
5  & 19.1 & 4.8 & 9.3 & 3.0 & 51.2 & 6.1 & 2.0 & 68.2    \\
10 & 14.7 & 2.5 & 7.0 & 0.9 & 52.1 & 5.5 & 0.7 & 62.5    \\
20 & 13.2 & 2.0 & 6.4 & 0.8 & 51.3 & 5.0 & 0.4 & 62.3    \\
30 & 12.5 & 2.0 & 6.1 & 0.6 & 51.4 & 4.9 & 0.4 & 60.8   \\
\specialrule{.1em}{.05em}{.05em} 
\end{tabular}
\label{tbl:validation}
\end{table}

\begin{table}[t]
\small
\caption{Segment error rate on the test set. Same notations are used as in Table~\ref{tbl:validation}.}
\label{tbl:test}
\begin{tabular}{c|P{0.4cm}P{0.4cm}|P{0.4cm}P{0.4cm}P{0.8cm}|P{0.4cm}P{0.4cm}P{0.8cm}}
\specialrule{.1em}{.05em}{.05em} 
\multirow{2}{*}{\begin{tabular}[c|]{@{}l@{}}\# \\  
\end{tabular}} & \multicolumn{2}{c|}{Cosine} & \multicolumn{3}{c|}{LP} & \multicolumn{3}{c}{GCN} \\               & mean         & std.        & mean  & std. & RER/\% & mean  & std.  & RER/\% \\ \hline
5  & 16.2 & 2.5 & 11.6 & 2.6 & 28.1 & 8.3 & 1.6 & 48.7 \\
10 & 13.8 & 1.5 & 10.1 & 1.4 & 26.9 & 7.4 & 0.5 & 46.6 \\
20 & 12.0 & 1.0 & 8.7  & 1.0 & 27.8 & 7.2 & 0.5 & 40.1 \\
30 & 11.6 & 0.7 & 8.4  & 0.6 & 27.6 & 7.4 & 0.5 & 36.4 \\
\specialrule{.1em}{.05em}{.05em} 
\end{tabular}
\end{table}

As expected, the mean error rate and stand derivation drops with more profile d-vectors provided. The GNN-based method outperforms label propagation significantly, especially with fewer profile d-vectors. However, the performance of the baseline method improves on the test set while the performance of graph based methods drop. This might be due to the discrepancy of acoustic characteristics between the \textit{dev} and \textit{test} sets, to which the baseline approach is less vulnerable since it is parameter free.

\section{Conclusion}
In this work, we applied graph-based semi-supervised learning methods for the speaker attribution task with speaker voice profiles. We build a graph of speech segments for each meeting with both the profile audio and meeting audio. We applied two methods to this task -- label propagation and a GNN-based method. Experiments on real multi-party meeting data showed that the graph-based methods outperformed the classifier-based methods significantly due to its use of meeting-wide structure information represented as graphs. Moving forward, we will extend the graph-based methods for online speaker identification and scenarios where some or all the speakers do not have voice profiles.  

\section{Acknowledgement}
The authors thank Tianyan Zhou and Yong Zhao for providing the d-vector extraction model, and Zhuo Chen for helpful discussions. Rudzicz is a CIFAR Chair in AI.

% \vfill\pagebreak

\bibliographystyle{IEEEtran}
% {\footnotesize \bibliography{refs}
\bibliography{refs}

\end{document}